\documentstyle[12pt,epsfig]{article}
\advance\oddsidemargin by -1.5cm
\advance\evensidemargin by -1cm
\def\be{\begin{equation}}
\def\ee{\end{equation}}

\begin{document}

~\vfill
\begin{center}
{\huge\bf Non-GUT Baryogenesis and Large Scale Structure of the Universe}
\vfill

D.P.Kirilova\footnote{Permanent address: Institute of Astronomy at
Bulgarian Academy of Sciences, Bul. Tsarigradsko Shosse 72, 1784 Sofia,
Bulgaria} 
and 
M.V.Chizhov\footnote{Permanent address: Centre of Space Research and
Technologies, Faculty of Physics, University of Sofia, 1164 Sofia,
Bulgaria, E-mail: mih@phys.uni-sofia.bg}\\ 
\ \\
{\it The Abdus Salam International Centre for Theoretical Physics,\\
Strada Costiera 11, 34014 Trieste, Italy}
\end{center}
\vfill
 
\begin{abstract}
    We discuss a mechanism for producing baryon density perturbations
during inflationary stage and study the evolution of the baryon 
charge density distribution
in the framework of the low temperature baryogenesis scenario. 
 This mechanism may be important for the 
 large scale structure formation of the Universe 
and particularly, may be essential for understanding the existence of 
a characteristic scale of $130h^{-1}$ Mpc (comoving size) in the 
distribution of the visible matter. 

The detailed analysis showed that both the observed very 
large scale of the visible matter distribution 
in the Universe and the observed baryon asymmetry value could 
naturally appear as a result of the evolution of a complex 
scalar field condensate, 
formed at the inflationary stage. 

Moreover, according to our model, the visible part of the Universe 
at present may consist of baryonic and antibaryonic regions, 
sufficiently separated, so that annihilation radiation is not observed.   
    
\end{abstract}
\vfill
keywords: cosmology -- large-scale structure - baryogenesis
\newpage
\section{Introduction} \indent

The large scale texture of the Universe shows a great complexity and 
variety of observed structures, it shows a strange pattern 
of filaments, voids and sheets. Moreover, due to the 
increasing amount of different types of observational data and 
theoretical analysis the last years,  it was realized, that there exists a 
characteristic very large scale of about $130h^{-1}$ Mpc in the 
large scale texture of the Universe. Namely, 
the galaxy deep pencil beam surveys (Broadhurst et al. 1988, 1990)
found an intriguing
periodicity in the very large scale distribution of the luminous
matter. The data consisted of several hundred redshifts
of galaxies, coming from four distinct surveys, in two narrow 
cylindrical volumes into the directions of the North and the South 
Galactic poles of our Galaxy, up to redshifts of more than $z\sim 0.3$,
combined to produce a well sampled distribution of galaxies by 
redshift on a linear scale extending to $2000h^{-1}$ Mpc.
The plot of the numbers
of galaxies as a function of redshifts displays a remarkably 
regular redshift distribution, with most galaxies lying in 
discrete peaks, with a periodicity over a scale of 
about $130h^{-1}$ Mpc comoving size. 

 It was realized also that   
the density peaks in the regular space distribution of galaxies in the 
redshift survey of Broadhurst et al. (1990), 
correspond to the location of superclusters, as defined by 
rich clusters of galaxies in the given direction (Bahcall 1991).    
The survey of samples in other 
directions, located near the South Galactic pole also gave indications 
for a regular distribution on slightly different scales near  
$100h^{-1}$ Mpc (Ettori et al. 1995, see also Tully et al. 1992 
and Guzzo et al. 1992, Willmer et al. 1994). 
This discovery of a large scale pattern at
the galactic poles was confirmed in a wider angle survey of 21 new pencil beams
distributed over 10 degree field at both galactic caps (Broadhurst et al. 1995)
and also by the new pencil-beam galaxy redshift data around the South Galactic
pole region (Ettori et al. 1997).

The analysis of other types of observations confirm the existence 
of this periodicity. Namely, such structure is consistent with the reported 
periodicity in the distribution of quasars and  radio galaxies 
(Foltz et al. 1989, Komberg et al. 1996, Quashnock et al. 1996, Petitjeau 1996, 
Cristiani 1998) 
and Lyman-$\alpha$ forest (Chu \& Zhu 1989);  the studies on 
spatial distribution of galaxies (both optical and IRAS) 
and clusters of galaxies (Kopylov et al. 1984, de Lapparent et al. 1986, 
Geller \& Hunchra 1989,  Hunchra et al. 1990, 
Bertshinger et al. 1990, Rowan-Robinson et al. 1990, 
Buryak et al. 1994, Bahcall 1992, Fetisova et al. 1993a,
Einasto et al. 1994, Cohen et al. 1996, Bellanger \& de Lapparent 1995) 
as well as peculiar velocity information (Lynden-Bell et al. 1988, 
Lauer \& Postman 1994, Hudson  et al. 1999)  
suggest the existence of a large scale superclusters-voids network 
with a characteristic scale around $130h^{-1}$ Mpc. 

An indication of the presence of this characteristic scale in the 
distribution of clusters has been found also from the studies of the 
correlation functions and power spectrum of clusters of galaxies 
(Kopylov et al. 1988, Bahcall 1991, Mo et al. 1992, Peacock \& West 1992, Deckel
et al. 1992, Saar et al. 1995, Einasto et al. 1993, Einasto \& Gramann 1993,
Fetisova et al. 1993b, Frisch et al. 1995, Einasto et al. 1997b,
Retzlaff et al. 1998, Tadros et al. 1997). The
galaxy correlation function of the Las Campanas
redshift survey also showed the presence of a secondary maximum at the same 
scale and a strong peak in the 2-dimensional power spectrum corresponding
to an excess power at about 100 Mpc (Landy et al. 1995, 1996, 
Shectman et al. 1996, Doroshkevich et al. 1996, 
 Geller at al. 1997, Tucker et al. 1999). 
The supercluster distribution
was shown also to be not random but rather  described as some weakly correlated 
network of superclusters and voids with typical mean separation 
of $100-150h^{-1}$ Mpc. Many known superclusters were identified with the 
vertices of an octahedron superstructure network (Battaner 1998). 
The network was proven to resemble a cubical lattice, with a periodic
distribution of the rich clusters along the main axis (coinciding with the
supergalactic $Y$ axis) of the network, with a step $\sim 130 h^{-1}$ 
Mpc (Toomet et al. 1999). These
results are consistent with the statistical analysis 
of the pencil beam surveys data (Kurki-Suonio et al. 1990, Amendola 1994),
which advocates a regular structure. 

Recently performed study of the whole-sky distribution of high density 
regions defined by very rich Abell and APM clusters of 
galaxies (Baugh 1996, Einasto et al. 1994, 1996, 1997a, 
Gaztanaga \& Baugh 1997, Landy et al. 1996, Retzlaff et al. 1998, 
Tadros et al. 1997, Kerscher 1998) confirmed from 
3-dimensional data the presence of the 
characteristic scale of about $130h^{-1}$ Mpc of the spatial 
inhomogeneity of the Universe, found by Broadhurst et al. (1988, 1990) 
from the one dimensional study. 
The combined evidence from cluster and CMB data 
(Baker et al. 1999, Scott et al. 1996) also favours the presence of a
peak at $130h^{-1}$ Mpc and a subsequent break in the initial power spectrum
(Atrio-Barandela et al. 1997, Broadhurst$\&$Jaffe 1999).
For a recent review of the regularity of the
Universe on large scales see Einasto (1997).
  
Concerning all these rather convenient data, pointing that different 
objects trace the same structure at large scales, we
are forced to believe 
in the real existence of the $130h^{-1}$ Mpc as a typical scale for 
the matter distribution in the Universe (see also Einasto et al 1998). 
However, this periodicity points to the 
existence of a significantly larger scale in the observed today
Universe  structure 
than predicted by standard models of structure formation by 
gravitational instability (Davis 1990 , Szalay et al. 1991, 
Davis et al. 1992, Luo \& Vishniac 1993, Bahcall 1994, 
Retzlaff et al. 1998, Atrio-Barandela et al. 1997, 
Lesgourgues et al. 1998, Meiksin et al. 1998, Eisenstein et al. 1998,
Eisenstein \& Hu 1997a, 1997b) and is rather to be regarded
as a new feature appearing only when very large scales ($>100
h^{-1}$ Mpc) are probed.

 The problem of the generation of the spatial periodicity in the density  
distribution of luminous matter at large scales was discussed 
in numerous publications (Lumsden et al. 1989,  Ostriker \& Strassler 1989,
Davis 1990, Coles 1990, Kurki-Suonio et al. 1990,  Trimble 1990, 
Kofman et al. 1990, 
Ikeuchi \& Turner 1991, van de Weygaert 1991,  Buchert \& Mo 1991,
SubbaRao \& Szalay 1992, Coleman \& Pietronero 1992,  Hill et al. 1989, 
Tully et al. 1992, Chincarini 1992, 
Weis \& Buchert 1993, Atrio-Barandela et al. 
1997, Lesgourgues et al. 1998, Eisenstein \& Hu 1997a, Meiksin et al. 1998,
Eisenstein et al. 1998, etc.). It was shown that a random structure 
could not explain the observed distribution. Statistical analysis 
of the deviations from periodicity showed that even for a perfectly 
regular structure a somewhat favoured direction and/or location 
within the structure may be required. The presence of the 
observed  periodicity  up to a great distance and in different 
directions seams rather amazing. Having in mind this results and 
 the difficulties that perturbative models encounter in 
explaining the very large scale structure formation (namely the 
existence of the very large characteristic scale and the 
periodicity of the visible matter distribution), we chose 
another way of exploration, namely, we assume these as a typical 
new feature characteristic only for very large scales ($>100
h^{-1}$ Mpc). I. e. we consider the possibility that 
density fluctuations required to explain the present cosmological 
largest scale structures of the universal texture may have arisen 
in a different from the standard way, they may be a 
result from a completely different mechanism not necessarily 
with gravitational origin. 

Such a successful mechanism was already proposed  
(Chizhov \& Dolgov 1992) and analyzed in the framework of 
high-temperature baryogenesis scenarios.\footnote{By high-temperature 
baryogenesis scenarios we denote here those 
scenarios which proceed at very high energies of the order of the Grand
Unification scale, and especially the GUT baryogenesis scenarios.  
In contrast, low temperature baryogenesis scenarios like Affleck and Dine
scenario and electroweak baryogenesis, proceed at several orders of
magnitude lower energies.} According to the discussed mechanism  an
additional 
complex scalar field (besides inflaton) is assumed to be present 
at the inflationary epoch, and it yields the extra power at the 
very large scale discussed.
Primordial baryonic fluctuations are produced during the inflationary
period, 
due to the specific evolution of the space distribution of the 
complex scalar field, carrying the baryon charge. 

In the present work we study the possibility of generating of 
 periodic space distribution of primordial baryon density fluctuations 
at the end of inflationary stage, applying 
this mechanism for the case of low temperature baryogenesis with 
baryon charge condensate of Dolgov \& Kirilova (1991). 
 The preliminary analysis of this problem, provided in Chizhov 
\& Kirilova (1994), proved its usefulness in that case.  
Here  we provide detail analysis of the evolution 
of the baryon density perturbations from the inflationary epoch 
till the baryogenesis epoch and describe the evolution of the 
spatial distribution of the baryon density. The production of matter-antimatter 
asymmetry in this scenario proceeds generally
at low energies ($\le 10^{9}$ GeV). This is of special
importance having in mind that the low-temperature baryogenesis scenarios 
are the preferred ones, as far as for their realization in the postinflationary
stage it is not necessary to provide considerable reheating temperature 
typical for GUT high temperature baryogenesis scenarios. 
Hence, the discussed model (Dolgov \& Kirilova 1991) has several attractive
features: 
(a) It is compatible with the inflationary models as far as it does not
suffer from the problem of insufficient reheating. (b) Generally, this
scenario evades the problem of washing out the previously produced 
baryon asymmetry at the
electroweak transition. (c) And as it will be proved in the following it may  
solve the problem of large scale periodicity of the visible matter.

  It was already discussed in 
(Dolgov 1992, Chizhov \& Dolgov 1992)
a periodic in space baryonic density distribution can be obtained
provided that the following assumptions are realized:

(a) There exists a complex scalar field $\phi$ with a mass small in
comparison with the Hubble parameter during inflation.

(b) Its potential contains nonharmonic terms.

(c) A condensate of $\phi$ forms during the inflationary stage and it
is a slowly varying function of space points.

All these requirements can be naturally fulfilled in our scenario of the
scalar field condensate baryogenesis (Dolgov \& Kirilova 1991)
and in low temperature baryogenesis scenarios based on the Affleck and Dine
mechanism (Affleck \& Dine 1985).

 In case when the potential of $\phi$ is not strictly harmonic 
the oscillation period
depends on the amplitude $P(\phi_0(r))$, and it on its turn depends
on $r$. Therefore, a
monotonic initial space distribution will soon result into spatial
oscillations of $\phi$ (Chizhov \& Dolgov 1992). 
 Correspondingly, the baryon charge, contained in $\phi$:
$N_B=i\phi^* \stackrel {\leftrightarrow} {\partial}_0 \phi$, will
have quasi-periodic behavior. During Universe expansion the
characteristic scale of the variation of $N_B$ will be inflated up to
a cosmologically interesting size. Then, if $\phi$ has not reached the
equilibrium point till the baryogenesis epoch $t_B$, the
baryogenesis would make a snapshot of the space distribution of
$\phi(r,t_B)$ and $N_B(r,t_B)$, and thus the
present periodic distribution of the visible matter may date from the
spatial distribution of the baryon charge contained in the $\phi$
field at the advent of the $B$-conservation epoch.

Density fluctuations with a comoving size today of $130 h^{-1}$ Mpc 
reentered the horizon at late times  at a redshift of about 10 000
and  a mass of $10^{18}M_o$. After recombination the Jeans mass 
becomes less than the horizon and the fluctuations of this large 
mass begin to grow. We propose that these baryonic fluctuations, 
periodically spaced, lead to an enhanced formation of galaxy superclusters
at the peaks of baryon overdensity.
The concentration of baryons into periodic shells may have catalysed 
also the clustering of matter coming from the inflaton decays onto 
these ``baryonic nuclei".
After baryogenesis proceeded, superclusters 
may have formed at the high peaks of the background field  
(the baryon charge carrying scalar field, we discuss). 
(See the results of the statistical analysis 
(Plionis 1995), confirming the idea that clusters formed at 
the high peaks of background field, which is analogous to our
assumption.) 
We imply that afterwards the 
self gravity mechanisms might have optimized the arrangement of 
this structure into the thin regularly spaced dense baryonic shells 
and voids in between with the characteristic size of $130 h^{-1} Mpc$ 
observed today. 

 The analysis showed that in the
framework of our scenario both the generation of the baryon asymmetry
and the periodic distribution of the baryon density can be explained
simultaneously as due to the evolution of a complex scalar field.

Moreover, for a certain range of parameters the model predicts that the
Universe may consist of sufficiently separated baryonic and antibaryonic
shells. This possibility was discussed in more detail
elsewhere (Kirilova 1998).
This is an interesting possibility as far as the observational 
data of antiparticles in cosmic rays and the gamma rays data do not 
rule out the possibility for existence of superclusters of galaxies 
of antimatter 
in the Universe (Steigman 1976, Ahlen et al. 1982, 1988,  Stecker
1985, 1989, Gao et al. 1990). 
The observations exclude the possibility of
noticeable amount of antimatter in our Galaxy, however, they are not
sensitive enough to test the existence of antimatter extragalactic
regions. I.e. current experiments 
(Salamon et al. 1990, Ahlen et al. 1994, Golden et al. 1994, 1996,
Yoshimura et al. 1995, Mitchell et al. 1996, 
Barbiellini \& Zalateu 1997, Moiseev et al. 1997, Boesio et al. 1997, 
Orito et al. 1999, etc.)
put only a lower limit  on the 
distance to the nearest antimatter-rich region, namely $\sim 20$
Mpc. Future searches for antimatter among cosmic rays are expected to
increase this lower bound by an order of magnitude. Namely, 
the reach of the AntiMatter Spectrometer is claimed to exceed 150 Mpc 
(Ahlen et al. 1982) and its sensitivity is three orders of magnitudes
better than that of the previous experiments (Battiston 1997, 
Plyaskin et al. 1998).
For a more detail discussion on the problem of existence of noticeable
amounts of antimatter at considerable distances see Dolgov (1993), 
Cohen et al. (1998), Kinney et al. (1997).

The following section describes the baryogenesis model and the last section
deals with the generation of the periodicity of the baryon density 
and discusses the results. 

\section{Description of the model. Main characteristics.} \indent

Our analysis was performed in the framework of the low temperature 
non-GUT baryogenesis model described in (Dolgov \&
Kirilova 1991), based on the Affleck and Dine SUSY GUT motivated 
mechanism for generation of the baryon asymmetry  
(Affleck \& Dine 1985). In this section we describe the main  
characteristics of the baryogenesis model, which are 
essential for the investigation of the periodicity 
in the next section. For more detail please see the original paper.

 \subsection{Generation of the baryon condensate.} \indent

     The essential ingredient of the model is a squark condensate 
$\phi$ with a nonzero baryon charge. It naturally appears in supersymmetric
theories and is a scalar superpartner of quarks.
The condensate $<\phi>\neq0$ is formed during the
inflationary period  as a result of
the enhancement of quantum fluctuations of the $\phi$ field 
(Vilenkin \& Ford 1982, Linde 1982, Bunch \& Davies 1978, Starobinsky 1982):
$<\phi^2>=H^3t/4\pi^2$. The baryon charge of the field is not
conserved at large values of the field amplitude due to the presence
of the B nonconserving self-interaction terms in the field's
potential. As a result, a condensate of a baryon charge (stored in $<\phi>$) 
is developed during inflation with 
 a baryon charge density of the order of $H^3_I$,
where $H_I$ is the Hubble parameter at the inflationary stage.

\subsection{Generation of the baryon asymmetry.} \indent

 After inflation $\phi$ starts to oscillate around its equilibrium
point with a decreasing amplitude. This decrease is due to the
Universe expansion and to the particle production by the oscillating
scalar field (Dolgov \& Kirilova 1990, 1991). Here we discuss the
simple case of particle production when $\phi$ decays into fermions 
and there is no parametric resonance. We expect that the case of decays
into bosons due to parametric resonance (Kofman et al. 1994, 1996, 
Shtanov et al. 1995, Boyanovski et al. 1995, Yoshimura 1995,
Kaiser 1996), especially in the broad resonance case, will lead to an
explosive decay of the condensate,  and hence an insufficient  baryon
asymmetry. Therefore, we explore the more promising case  of $\phi$ 
decaying into fermions.  

In the expanding Universe $\phi$
satisfies the equation
\be
\ddot{\phi}-a^{-2}\partial^2_i\phi+3H\dot{\phi}+
{1 \over 4}\Gamma\dot{\phi}+U'_{\phi}=0,
\ee
where $a(t)$ is the scale factor and $H=\dot{a}/a$.

The potential $U(\phi)$ is chosen in the form
\be
U(\phi)={\lambda_1\over 2}|\phi|^4
+{\lambda_2\over 4}(\phi^4+\phi^{*4})+{\lambda_3\over 4}|\phi|^2
(\phi^2+\phi^{*2})
\ee
The mass parameters of the potential are assumed small in comparison
to the Hubble constant during inflation $m \ll H_I$. In
supersymmetric theories the constants $\lambda_i$ are of the order of
the gauge coupling constant $\alpha$. A natural value of $m$ is
$10^2\div10^4$ GeV. The initial values for the field variables can be
derived from the natural assumption that the energy density of $\phi$
at the inflationary stage is of the order $H^4_I$, then
$\phi^{max}_o \sim H_I\lambda^{-1/4}$ and $\dot{\phi_o}=0$.

The term $\Gamma\dot{\phi}$ in the equations of motion explicitly
accounts for the eventual damping of $\phi$ as a result of particle
creation processes. 
The explicit account 
for the effect of particle creation processes in the equations of motion
was first provided in (Chizhov \& Kirilova 1994, 
Kirilova \& Chizhov 1996).
The production rate $\Gamma$ was calculated in
(Dolgov \& Kirilova 1990). 
For simplicity here  we have used the perturbation
theory approximation for the production rate $\Gamma = \alpha \Omega$, where
$\Omega$ is the frequency of the scalar field.\footnote{For the toy model, 
we discuss here, we consider this approximation instructive enough.}  
For $g<\lambda^{3/4}$,  $\Gamma$ considerably exceeds the rate of the
ordinary decay of the field $\Gamma_m=\alpha m$.  
Fast oscillations of $\phi$ after inflation
result in particle creation due to the coupling of the scalar field
to fermions $g\phi \bar{f}_1 f_2$, where $g^2/4\pi = \alpha_{SUSY}$.
Therefore, the amplitude of $\phi$ is damped as $\phi \rightarrow
\phi \exp(-\Gamma t/4)$ and the baryon charge, contained in the
$\phi$ condensate, is considerably  reduced. It was discussed in 
detail in Dolgov \& Kirilova (1991) that for a constant $\Gamma$ this 
reduction is exponential and generally, for a natural range of the
model's parameters, the baryon asymmetry is waved away till 
baryogenesis epoch as a result of the particle creation processes. 
Fortunately, in the case without flat directions of the potential, 
the production rate is a decreasing function of time, so that the  
damping process may be slow enough for a considerable range of 
acceptable model parameters values
of $m$, $H$, $\alpha$, and $\lambda$, so that the baryon charge
contained in $\phi$ may survive until the advent of the
$B$-conservation epoch. Generally, in cases of more effective particle
creation, like in the case with flat directions in the potential, 
or in the case  when  $\phi$ decays spontaneously into bosons due to
parametric resonance, the discussed mechanism of the baryon asymmetry 
generation cannot be successful. Hence, 
it cannot be useful also for the generation of the matter periodicity. 

\subsection{Baryogenesis epoch $t_B$.} \indent

   When inflation is over and $\phi$ relaxes to its equilibrium state, its
coherent oscillations produce an excess of quarks over antiquarks 
(or vice versa) depending on the initial sign of the baryon charge 
condensate. This charge, diluted further by some entropy
generating processes, dictates the observed baryon asymmetry.
This epoch when  $\phi$ decays to quarks with non-zero
average baryon charge and thus induces baryon asymmetry we call
baryogenesis epoch. 
The baryogenesis epoch $t_B$ for our model  coincides with the
advent of the baryon conservation epoch, i.e. the time after
which the mass terms in the equations of motion cannot be neglected.  
In the original version (Affleck \& Dine 1985) this epoch corresponds to
energies 
$10^{2}-10^{4}$ GeV. However, as it was already explained, 
the amplitude of $\phi$ may be reduced much more quickly due to the 
particle creation processes and as  a result, depending on the   
model's parameters the advent of
this  epoch may be considerably earlier.
For the correct estimation of  $t_B$ and the
value of the generated baryon asymmetry, it
is essential to account for the eventual damping of the field's
amplitude due to particle production processes by an external
time-dependent scalar field, which could lead to a strong reduction
of the baryon charge contained in the condensate.

\section{Generation of the baryon density periodicity.} \indent

      In order to explore the spatial distribution behavior of the
scalar field and its evolution during Universe expansion it is 
necessary to analyze eq.(1). We have
made the natural assumption that initially $\phi$ is a slowly 
varying function of the space coordinates $\phi(r,t)$. 
 The space derivative term can be safely neglected because of the
exponential rising of the scale factor $a(t)\sim\exp(H_It)$. Then
the equations of motion for $\phi=x+iy$ read
\begin{eqnarray}
&&\ddot{x}+3H\dot{x}+{1 \over 4} \Gamma_x \dot{x}+
(\lambda+\lambda_3)x^3+\lambda'xy^2=0 
\nonumber
\\
&&\ddot{y}+3H\dot{y}+{1 \over 4} \Gamma_y \dot{y}+
(\lambda-\lambda_3)y^3+\lambda'yx^2=0
\end{eqnarray}
where $\lambda=\lambda_1+\lambda_2$, $\lambda'=\lambda_1-3\lambda_2$.

In case when at the end of inflation the Universe is dominated by a
coherent oscillations of the inflaton field
$\psi=m_{PL}(3\pi)^{-1/2}\sin(m_{\psi}t)$, the Hubble parameter is
$H=2/(3t)$. In this case it is convenient to make the substitutions
$x=H_I(t_i/t)^{2/3}u(\eta)$, $y=H_I(t_i/t)^{2/3}v(\eta)$ where
$\eta=2(t/t_i)^{1/3}$. The functions $u(\eta)$ and $v(\eta)$ satisfy
the equations
\be
\begin{array}{c}
u''+ 0.75\; \alpha\Omega_u (u'-2u\eta^{-1})+
u[(\lambda+\lambda_3) u^2+\lambda'v^2-2\eta^{-2}]=0\\
v''+ 0.75\; \alpha\Omega_v (v'-2v\eta^{-1})+
v[(\lambda-\lambda_3) v^2+\lambda'u^2-2\eta^{-2}]=0.
\end{array}
\ee
The baryon charge in the comoving volume $V=V_i(t/t_i)^2$ is
$B=N_B \cdot V=2 (u'v-v'u)$.
 The numerical calculations were performed for 
$u_o,v_o \in [0, \lambda^{-1/4}]$,
$u'_o,v'_o \in [0, 2/3 \lambda^{-1/4}]$.
For simplicity we considered the case: 
$\lambda_1 > \lambda_2 \sim \lambda_3$,
when the unharmonic oscillators $u$ and $v$ are weakly coupled. For
each set of parameter values of the model $\lambda_i$ we have
numerically calculated the baryon charge evolution $B(\eta)$ for
different initial conditions of the field corresponding to the
accepted initial monotonic space distribution of the field (see
Figs. 1,2). 
 
The numerical analysis confirmed 
the important role of particle creation processes for 
baryogenesis models and large scale 
structure periodicity (Chizhov \& Kirilova 1994, 1996) 
which were obtained from an approximate analytical 
solution. In the present work  we have accounted for
particle creation processes explicitly.
\footnote{It was shown, that the damping effect due to the 
particle creation is proportional to the initial amplitudes of the field.
 As far as the particle creation rate is proportional to
the field's frequency, it can be concluded that the frequency depends
on the initial amplitudes. This result confirms our  analytical estimation 
provided in earlier works.} 

 The space distribution of the baryon charge is calculated for the moment
$t_B$. It is obtained from the evolution analysis
$B(\eta)$ for different initial values of the field, corresponding to
its initial space distribution $\phi(t_i,r)$ (Fig. 3). As it was
expected, in the case of nonharmonic field's potential, the initially
monotonic space behavior is quickly replaced by space oscillations of
$\phi$, because of the dependence of the period on the amplitude,
which on its turn is a function of $r$. As a result in different
points different periods are observed and space behavior of $\phi$
becomes quasiperiodic. 
Correspondingly, the space
distribution of the baryon charge contained in $\phi$ becomes
quasiperiodic as well. Therefore, the space distribution of baryons
at the moment of baryogenesis is found to be periodic. 

The observed 
space distribution of the visible matter
today is defined by the space distribution of the baryon charge of
the field $\phi$ at the moment of baryogenesis $t_B$, $B(t_B,r)$.
So, that at present the visible part of the
Universe consists of baryonic shells, divided by vast underdense regions.
{\it For a wide range of parameters' values the observed average distance
of $130h^{-1}$ Mpc between matter shells in the Universe can be obtained.
The parameters of the model ensuring the necessary observable size
between the matter domains belong to the range of parameters for
which the generation of the observed value of the baryon asymmetry
may be possible in the model of scalar field condensate baryogenesis.}
This is an attractive feature of this model because both
the baryogenesis and the large scale structure periodicity of the
Universe can be explained simply through the evolution of a single
scalar field.

Moreover, for some model's 
variations the presence of 
vast antibaryonic regions in the Universe is predicted.   
This is an interesting possibility as far as the observational data
do not rule out the possibility of antimatter superclusters in the 
Universe.
 The  model proposes an elegant mechanism for
achieving a sufficient  separation between regions occupied 
by baryons and those occupied by antibaryons, necessary in order 
to inhibit the contact of matter and antimatter regions with considerable 
density.  

It is interesting, having in mind the
positive results of this investigation, to provide a more precise study
of the question for different possibilities of particle creation and
their relevance for the discussed scenario of baryogenesis and
periodicity generation.  In the case of narrow-band resonance
decay the final state interactions regulate the decay rate,
parametric amplification is effectively suppressed 
(Allahverdi \& Campbell 1997) and does not drastically enhance the decay rate.
Therefore, we expect that this case will be interesting to explore.
Another interesting case may be the case of strong dissipative processes
of the products of the parametric resonance. As far as the dissipation
reduces the resonant decay rate (Kolb et al. 1996,
Kasuya \& Kawasaki 1996) it may be worthwhile to consider such a model as
well.

\section{Acknowledgments} \indent

We are glad to thank A.D.Dolgov for stimulating our interest in 
this problem. 
We are thankful to ICTP, Trieste, where this work 
was finished, for the financial support and hospitality. 
We are grateful also to the referee for the useful remarks and suggestions.

This work was partially financially supported by Grant-in-Aid for Scientific 
Research F-553 from the Bulgarian Ministry of 
Education, Science and Culture.\\

\ \\

\pagebreak[3]
{\bf \Large References}\\

\noindent Affleck, I. \& Dine, M., 1985, Nucl. Phys. B, 249, 361

\noindent Ahlen, S. et al., 1982, Ap.J., 260, 20

\noindent Ahlen, S. et al., 1988, Phys. Rev. Lett., 61, 145

\noindent Ahlen, S. et al., 1994, N.I.M. A, 350, 351

\noindent Allahverdi, R. \& Campbell, B., 1997, Phys. Lett. B, 395, 169

\noindent Amendola, L., 1994, Ap.J., 430, L9

\noindent Atrio-Barandela, F. et al., 1997, JETP Lett., 66, 397

\noindent Bahcall, N., 1991, Ap.J., 376, 43

\noindent Bahcall, N., 1992, in Clusters and Superclusters of Galaxies,
Math. Phys. Sciences, 366                                                   

\noindent Bahcall, N., 1994, Princeton observatory preprints, POP-607

\noindent Baker, J. et al., 1999, astro-ph/9904415

\noindent Barbiellini, G. \& Zalateu, M., 1997, INFN-AE-97-29 

\noindent Battaner, E., 1998, A$\&$Ap, 334, 770

\noindent Battiston, R., 1997, hep-ex/9708039

\noindent Baugh, C., 1996, MNRAS, 282, 1413 

\noindent Bellanger, C. \& de Lapparent, V., 1995, Ap.J., 455, L103 

\noindent Bertshinger, E., Deckel, A. \& Faber, S., 1990, Ap.J., 364, 370

\noindent Boesio, M. et al., 1997, Ap.J., 487, 415

\noindent Boyanovski, D. et al., 1995, Phys. Rev. D, 51, 4419

\noindent Broadhurst, T.J., Ellis, R.S. \& Shanks, T., 1988, MNRAS, 235,
827

\noindent Broadhurst, T.J., Ellis, R.S., Koo, D.C. \& Szalay, A.S.,
1990, Nature, 343, 726

\noindent Broadhurst, T.J. et al., 1995, in Wide Field Spectroscopy and 
the Distant Universe, eds. Maddox, S. and Aragon-Salamanca, A., 178

\noindent Broadhurst, T. \& Jaffe A., 1999, astro-ph/9904348

\noindent Buchert, T. \& Mo, H., 1991, A\&Ap, 249, 307

\noindent Bunch, T.S. \& Davies, P.C.W., 1978, Proc. R. Soc. London A, 360, 117

\noindent Buryak, O. E., Doroshkevich, A.G. \& Fong, R., 1994, Ap.J., 434, 24

\noindent Chincarini, G., 1992, in Clusters and Superclusters of Galaxies, 
ed. Fabian, A., Series C, Math. Phys. Sci. 366, 253

\noindent Chizhov, M.V. \& Dolgov, A.D., 1992, Nucl. Phys. B, 372, 521

\noindent Chizhov, M.V. \& Kirilova, D.P., 1994, JINR Comm. E2-94-258, Dubna

\noindent Chizhov, M.V. \& Kirilova, D.P., 1996, Ap\&A Tr, 10, 69

\noindent Chu, Y. \& Zhu, X., 1989, A\&A, 222, 1

\noindent Cohen, J. et al. 1996, Ap.J., 462, L9

\noindent Cohen, A., DeRujula, A. \& Glashow, S., 1998, Ap.J., 495, 539

\noindent Coleman, P. \& Pietronero, L., 1992, Phys. Rep., 213, 313

\noindent Coles, P., 1990, Nature, 346, 446

\noindent Cristiani, S., 1998, astro-ph/9811475, 
review at the MPA/ESO Cosmology Conference ``Evolution of Large-Scale 
Structure: From Recombination to Garching", Garching, 2-7 August

\noindent Davis, M., 1990, Nat, 343, 699 

\noindent Davis, M., Efstathiou, G., Frenk, C. \& White, S.,
 1992, Nature, 356, 489 

\noindent de Lapparent, V., Geller, M. \& Huchra, J., 1986, Ap.J., 302, L1

\noindent Deckel, A., Blumental, G.R., Primack, J.R. \& Stanhill, D., 1992, 
MNRAS, 257, 715  

\noindent Dolgov, A.D., 1992, Phys. Rep., 222, 311

\noindent Dolgov, A.D., 1993, Hyperfine Interactions, 76, 17

\noindent Dolgov, A.D. \& Kirilova, D.P., 1990, Yad. Fiz., 51, 273

\noindent Dolgov, A.D. \& Kirilova, D.P., 1991, J. Moscow Phys. Soc., 1, 217

\noindent Doroshkevich, A. et al., 1996, MNRAS, 283, 1281

\noindent Einasto, J. \& Gramann, M., 1993, Ap.J., 407, 443

\noindent Einasto, J., Gramann, M., Saar, E. \& Tago, E., 1993, MNRAS, 260, 705

\noindent Einasto, M. et al., 1994, MNRAS, 269, 301

\noindent Einasto, M. et al., 1996, astro-ph/9610088

\noindent Einasto, J., 1997, astro-ph/9711318, astro-ph/9711321

\noindent Einasto, J. et al., 1997a, MNRAS, 289, 801; A.Ap.Suppl., 123, 119;
Nature, 385, 139

\noindent Einasto, J. et al., 1997b, astro-ph/9704127, astro-ph/9704129

\noindent Einasto, J., 1998, astro-ph/9811432

\noindent Eisenstein, D. \& Hu, W., 1997a, astro-ph/9710252

\noindent Eisenstein, D. \& Hu, W., 1997b, astro-ph/9709112, astro-ph/9710303

\noindent Eisenstein, D. et al., 1998, Ap.J., 494, 1

\noindent Ettori, S., Guzzo, L. \& Tarenghi, M., 1995, MNRAS, 276, 689

\noindent Ettori, S., Guzzo, L. \& Tarenghi, M., 1997, MNRAS, 285, 218

\noindent Fetisova, T., Kuznetsov, D., Lipovetski, V., Starobinsky, A. 
\& Olowin, P., 1993a, Pisma v Astr. Zh., 19, 508; 

\noindent Fetisova, T., Kuznetsov, D., Lipovetski, V., Starobinsky, A. 
\& Olowin, P., 1993b, Astron. Lett., 19, 198

\noindent Foltz, C. et al., 1989, A.J., 98, 1959

\noindent Frisch, P. et al., 1995, A\&A 269, 611

\noindent Gao, Y. et al., 1990, Ap.J., 361, L37

\noindent Gaztanaga, E. \& Baugh, C., 1997, MNRAS in press, astro-ph/9704246

\noindent Geller, M. \& Hunchra, J., 1989, Science, 246, 897

\noindent Geller, M. et al., 1997, A.J., 114, 2205

\noindent Golden, R. et al., 1994, Ap.J., 436, 769

\noindent Golden, R. et al., 1996, Ap.J., 457, L103

\noindent Guzzo, L. et al., 1992, Ap.J., 393, L5

\noindent Hill, C., Fry, J. \& Schramm, D., 1989, 
Comm. Nucl. Part. Phys., 19, 25

\noindent Hudson, J. et al., 1999, astro-ph/9901001

\noindent Hunchra, J., Henry, J., Postman, M. \& Geller, M., 1990, 
Ap.J., 365, 66

\noindent Ikeuchi, S. \& Turner, E., 1991, MNRAS, 250, 519

\noindent Kaiser, D., 1996, Phys. Rev. D, 53, 1776

\noindent Kasuya, S. \& Kawasaki, M., 1996, Phys. Lett. B, 388, 686

\noindent Kerscher, M., 1998, astro-ph/9805088, astro-ph/9710207

\noindent Kinney, W. H., Kolb, E. W. \& Turner, M. S., 1997, 
Phys. Rev. Lett. 79, 2620

\noindent Kirilova, D. \& Chizhov, M., 1996, in Proc. Scient. Session 
 100th Anniversary of the Sofia University Astronomical Observatory, 
 Sofia, Naturela publ., Sofia, 114

\noindent Kirilova, D., 1998, Astron. Astrophys. Transections,  
 3, 211 (See also the updated version ICTP preprint IC/98/71, 1998) 

\noindent Kofman, L., Pogosyan, D. \& Shandarin, S., 1990, MNRAS, 242, 200

\noindent Kofman, L., Linde, A. \& Starobinsky, A., 1994, 
Phys. Rev. Lett., 73, 3195

\noindent Kofman, L., Linde, A. \& Starobinsky, A., 1996,
Phys. Rev. Lett., 76, 1011

\noindent Kolb, E., Linde, A. \& Riotto, A., 1996, 
Phys. Rev. Lett., 77, 4290

\noindent Komberg, B., Kravtsov, A. \& Lukash, V., 1996, MNRAS, 282, 713

\noindent Kopylov, A., Kuznetsov, D., Fetisova, T. \& Shvartzman, V., 1984, 
Astr. Zirk. 1347, 1 

\noindent Kopylov, A., Kuznetsov, D., Fetisova, T. \& Shvartzman, V., 1988, 
in Large Scale Structure in the Universe, eds. Audouze, J., 
Pelletan, M.-C. \& Szalay, A., 129

\noindent Kurki-Suonio, H., Mathews, G. \& Fuller, G., 1990,
 Ap.J., 356, L5                                                     

\noindent Landy, S. et al., 1995, Bull. American Astron. Soc., 187

\noindent Landy, S. et al., 1996, Ap.J., 456, L1

\noindent Lauer, T. \& Postman, M., 1994, Ap.J., 425, 418

\noindent Lesgourgues, J., Polarski, D. \& Starobinsky, A., 1998, 
MNRAS, 297, 769

\noindent Linde, A.D., 1982, Phys. Lett. B, 116, 335

\noindent Lumsden, S., Heavens, A. \& Peacock, J., 1989, MNRAS 238, 293

\noindent Luo, S. \& Vishniac, E.T., 1993, Ap.J., 415, 450

\noindent Lynden-Bell, D., Faber, S., Burstein, D., Davis, R.,
 Dressler, A., Terlevich, R. \& Wegner, G., 1988, Ap.J., 326, 19        

\noindent Meiksin, A., White, M. \& Peacock, J.A., 1998, astro-ph/9812214

\noindent Mitchell, J. et al., 1996,  Phys. Rev. Lett., 76, 3057

\noindent Mo, H. et al., 1992, A\&A, 257, 1; A\&A, 256, L23

\noindent Moiseev, A. et al., 1997, Ap.J., 474, 479

\noindent Orito, S. et al., 1999, astro-ph/9906426

\noindent Ostriker, J. \& Strassler, M., 1989, Ap.J., 338, 579

\noindent Peacock, J. \& West, M., 1992, MNRAS, 259, 494

\noindent Petitjeau, P., 1996, contributed talk at the ESO Workshop on 
``The Early Universe with the VLT" 1-4 April 

\noindent Plionis, M., 1995, MNRAS, 272, 869

\noindent Plyaskin, V. et al., AMS Collaboration, 1998, Surveys High
Energy Phys., 13, 177

\noindent Quashnock, J. et al., 1996, Ap.J.,  472, L69

\noindent Retzlaff, J. et al., 1998, astro-ph/9709044 v2 
 
\noindent Rowan-Robinson, R. et al., 1990, MNRAS, 247, 1 

\noindent Salamon, M. et al., 1990, Ap.J., 349, 78

\noindent Saar, V., Tago, E., Einasto, J., Einasto, M. \& Andernach, H.,
1995, astro-ph/9505053, in Proc. of XXX Moriond meeting

\noindent Scott, P. et al., 1996, Ap.J., 461, L1

\noindent Shectman, S. et al., 1996, Phys. Lett. B, 470, 172

\noindent Shtanov, Y., Traschen, J. \& Brandenberger, R., 1995,
Phys. Rev. D, 51, 5438

\noindent Starobinsky, A.A., 1982, Phys. Lett. B, 117, 175

\noindent Stecker, F., 1985, Nucl. Phys. B, 252, 25
 
\noindent Stecker, F., 1989, Nucl. Phys. B (Proc. Suppl.), 10, 93

\noindent Steigman, G., 1976, Ann. Rev. Astr. Ap., 14, 339

\noindent SubbaRao, M. U. \& Szalay, A.S., 1992, Ap.J., 391, 483

\noindent Szalay, A.S., Ellis, R.S., Koo, D.C. \& Broadhurst, T.J.,
1991, in proc. After the First Three Minutes, eds. Holt, S., 
Bennett, C. \& Trimble, V., New York: AIP, 261

\noindent Tadros, H., Efstathiou, G. \& Dalton, G., 1997, 
astro-ph/9708259, submitted to MNRAS

\noindent Toomet, O. et al., astro-ph/9907238

\noindent Trimble, V., 1990, Nature, 345, 665

\noindent Tucker, D., Lin, H. \& Shectman, S., 1999, astro-ph/9902023

\noindent Tully, R.B., Scaramella, R., Vetollani, G. \& Zamorani, G.,
1992, Ap.J., 388, 9

\noindent van de Weygaert, R., 1991, MNRAS, 249, 159

\noindent Vilenkin, A. \& Ford, L.H., 1982, Phys. Rev. D, 26, 1231

\noindent Weis, A. \& Buchert, T., 1993, A$\&$ Ap, 274, 1

\noindent Willmer, C. et al., 1994, Ap. J., 437, 560 

\noindent Yoshimura, M., 1995, Prog. Theor. Phys., 94, 873

\noindent Yoshimura, K. et al., 1995, Phys. Rev. Let., 75, 379

\newpage

\begin{center}{\Large Captions}
\end{center}
\ \\

{\bf Figure 1}: The evolution of the baryon charge $B(\eta)$ contained
in the condensate $<\phi>$ for $\lambda_1=5 \times 10^{-2}$,
$\lambda_2=\lambda_3=\alpha=10^{-3}$, $H_I/m=10^7$,
$\phi_o=H_I\lambda^{-1/4}$, and $\dot{\phi}_o=0$.

{\bf Figure 2}: The evolution of the baryon charge $B(\eta)$ contained
in the condensate $<\phi>$ for $\lambda_1= 5 \times 10^{-2}$,
$\lambda_2=\lambda_3=\alpha=10^{-3}$, $H_I/m=10^7$,
$\phi_o={1 \over 50}H_I\lambda^{-1/4}$, and $\dot{\phi}_o=0$.

{\bf Figure 3}: The space distribution of baryon charge at the moment
of baryogenesis for $\lambda_1=5 \times 10^{-2}$, $\lambda_2=\lambda_3=
\alpha=10^{-3}$, $H_I/m=10^7$.

\newpage

\begin{figure}
\epsfig{file=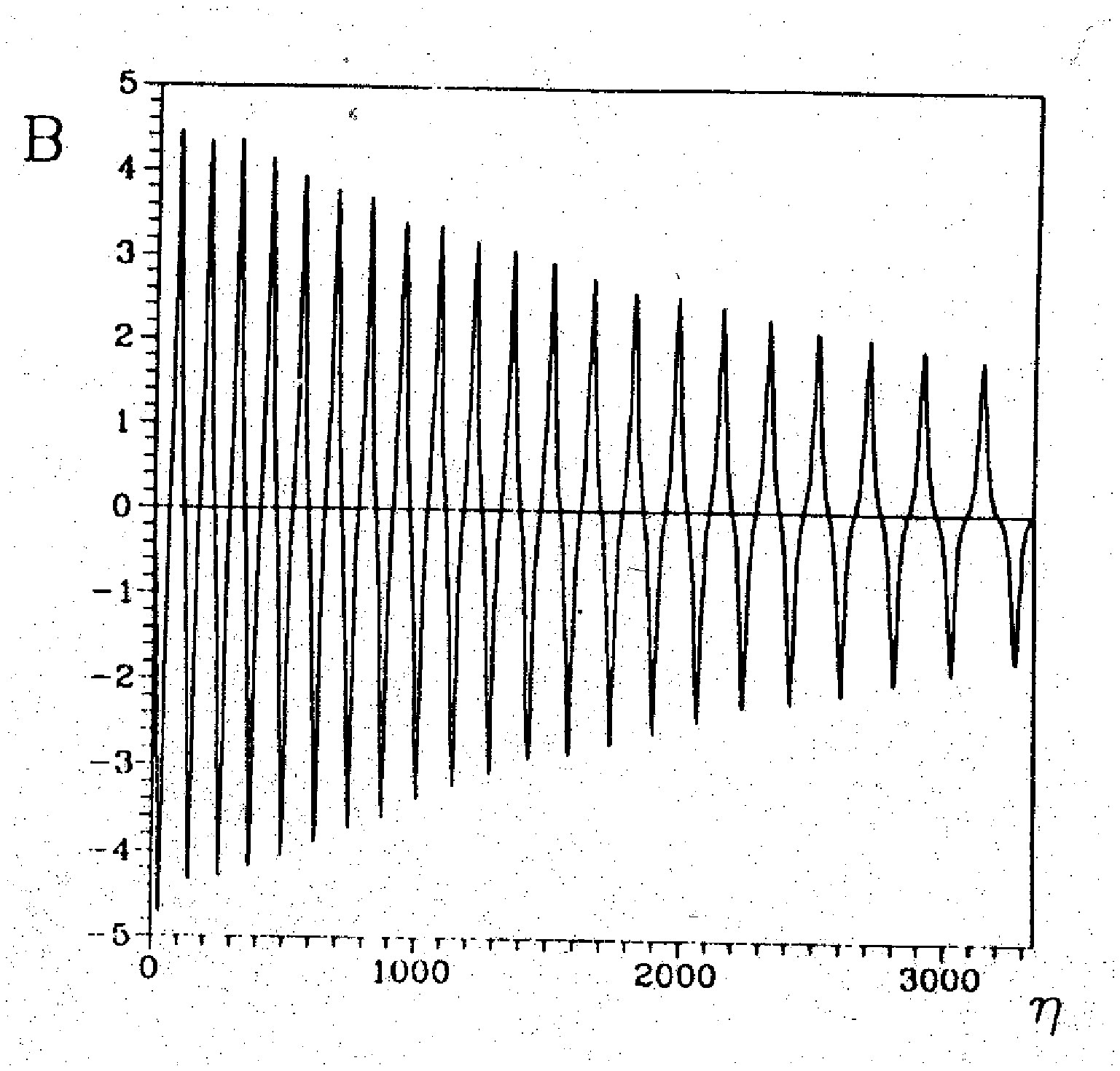,height=10cm,width=13cm}
\caption{The evolution of the baryon charge $B(\eta)$ contained
in the condensate $<\phi>$ for $\lambda_1=5 \times 10^{-2}$,
$\lambda_2=\lambda_3=\alpha=10^{-3}$, $H_I/m=10^7$,
$\phi_o=H_I\lambda^{-1/4}$, and $\dot{\phi}_o=0$.}
\end{figure}

\begin{figure}
\epsfig{file=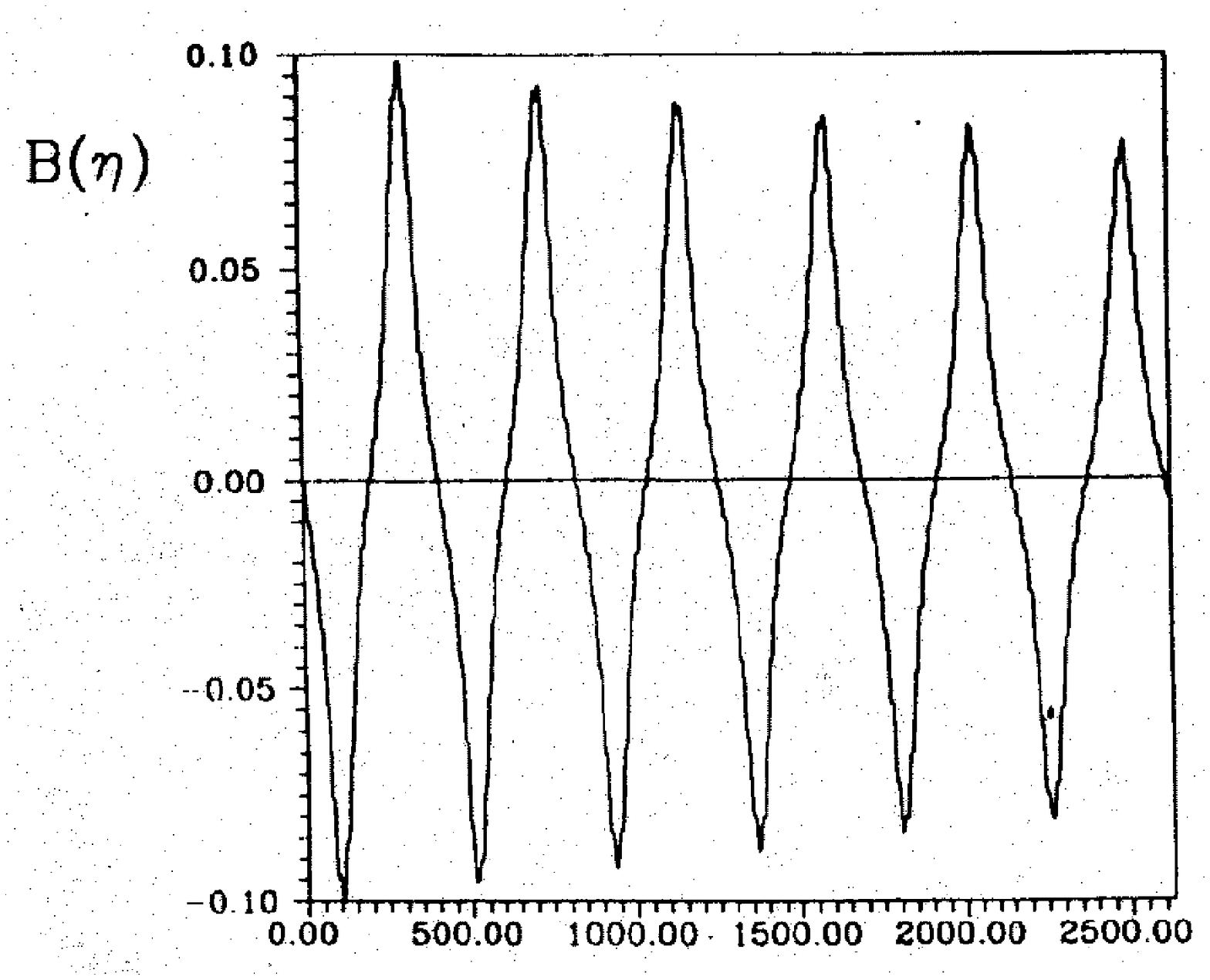,height=10cm,width=13cm}
\caption{The evolution of the baryon charge $B(\eta)$ contained
in the condensate $<\phi>$ for $\lambda_1= 5 \times 10^{-2}$,
$\lambda_2=\lambda_3=\alpha=10^{-3}$, $H_I/m=10^7$,
$\phi_o={1 \over 50}H_I\lambda^{-1/4}$, and $\dot{\phi}_o=0$.}
\end{figure}

\begin{figure}
\epsfig{file=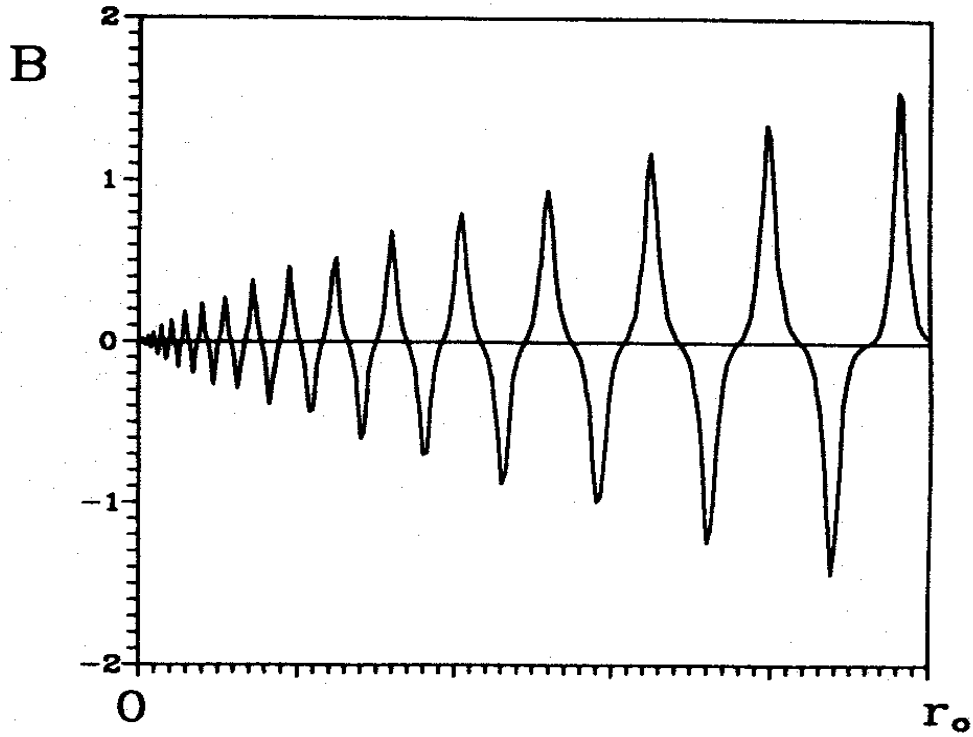,height=10cm,width=13cm}
\caption{The space distribution of baryon charge at the moment
of baryogenesis for $\lambda_1=5 \times 10^{-2}$, $\lambda_2=\lambda_3=
\alpha=10^{-3}$, $H_I/m=10^7$.}
\end{figure}

\end{document}